\documentclass[12pt]{iopart}
\usepackage{amssymb}
\usepackage{graphicx,bm}
\usepackage{setstack, cite}
\usepackage{lscape}
\begin{document}
\title[]{\bf Exact quantization of a PT symmetric (reversible) Li\'{e}nard type nonlinear
oscillator}

\author{V~Chithiika Ruby, M~Senthilvelan and M. Lakshmanan}

\address{Centre for Nonlinear Dynamics, School of Physics,
Bharathidasan University, Tiruchirapalli - 620 024, India.}

\begin{abstract}
We carry out an exact quantization of a PT symmetric (reversible) Li\'{e}nard type one dimensional
nonlinear oscillator both semiclassically and quantum mechanically. The associated
time independent classical Hamiltonian is of non-standard type and is invariant under a
combined coordinate reflection and time reversal transformation. We use von Roos
symmetric ordering procedure to write down the appropriate quantum Hamiltonian. While
the quantum problem cannot be tackled in coordinate space, we show how the problem
can be successfully solved in momentum space by solving the underlying Schr\"{o}dinger
equation therein.  We obtain explicitly the eigenvalues and eigenfunctions (in momentum space) and deduce
the remarkable result that the spectrum agrees exactly with that of the linear harmonic
oscillator, which is also confirmed by a semiclassical modified Bohr-Sommerfeld quantization rule,
while the eigenfunctions are completely different.
\end{abstract}

\maketitle

\section{Introduction}
In a previous paper \cite{chand1},  Chandrasekar and two of the present
authors have presented a conservative description for the Li\'{e}nard type
one dimensional nonlinear oscillator, namely
\begin{equation}
\qquad \ddot{x} + k x \dot{x} + \frac{k^2}{9} x^{3} + \omega^2 x = 0,
\label{mee}
\end{equation}
where overdot denotes differentiation with respect to $t$ and $k$ and $\omega^2$ are real
parameters.  Expressing (\ref{mee}) as a sysytem of first order
equations, $\dot{x} = y \equiv F_{1}(x, y), \; \dot{y} = - k x y - \frac{k^2}{9} x^3 - \omega^2 x \equiv F_2(x, y)$, one
can note that the divergence of the flow function $\vec{F} = F_1\vec{i} + F_2 \vec{j}$ of (\ref{mee}) is
non-zero $(\vec{\nabla} . \vec{F} = \frac{\partial F_1}{\partial x} + \frac{\partial F_2}{\partial y}= - k x)$.
Also equation (\ref{mee}) is invariant under the PT or reversible transformation,
$x \rightarrow -x$ and $t \rightarrow -t$ \cite{bend}.  In spite of these, system (\ref{mee})
admits a time independent Hamiltonian of the form
\begin{eqnarray}
\fl \;\;\; H(x, p) = \frac{9 \omega^4}{2 k^2}\left[2 - \frac{2 k }{3 \omega^2} p - 2 \left(1 - \frac{2 k}{3 \omega^2} p\right)^{\frac{1}{2}}+ \frac{k^2 x^2}{9\omega^2}\left(1 - \frac{2 k}{3 \omega^2} p\right)\right], \; -\infty < p \le \frac{3\omega^2}{2 k}, \qquad
\label{ham1}
\end{eqnarray}
which is of non-standard type, that is the coordinates and potentials
are mixed so that the Hamiltonian cannot be written as just the sum of the kinetic
and potential energy terms alone, including velocity dependent terms. The corresponding Lagrangian $L$ is given by
\begin{eqnarray}
\fl \qquad \qquad \; \; L =  \frac{27 \omega^6}{2 k^2}\left(\frac{1}{k\dot{x} + \frac{k^2}{3} x^2 + 3\omega^2}\right) + \frac{3\omega^2}{2 k} \dot{x} - \frac{9 \omega^4}{2 k^2},
\end{eqnarray}
and the conjugate momentum is
\begin{eqnarray}
\fl \qquad \qquad \; \; p = \frac{\partial L}{\partial \dot{x}} = - \frac{27 \omega^6}{2 k(k \dot{x}+\frac{k^2}{3} x^2 + 3 \omega^2)^2} + \frac{3\omega^2}{2 k}.
\label{mom}
\end{eqnarray}

We note here that the system (\ref{mee}) also admits an alternate Lagrangian/ Hamiltonian
(see for example Ref. \cite{chand1}). However we consider the Hamiltonian given in the
form (\ref{ham1}) only since as $k \rightarrow 0$ the Hamiltonian (\ref{ham1}) reduces to
the linear harmonic oscillator Hamiltonian, as the equation (\ref{mee}) does. We mention
here that the parameters $k$ and $\omega^2$ can be rescaled with appropriate scaling in $x$
and $t$. However,  to describe the physical properties of
this system in the classical, semi-classical and quantum levels, we retain the parameters
$k$ and $\omega^2$ and do not scale them away. We also note here that the Hamiltonian (\ref{ham1}) with
the definition of $p$ given by (\ref{mom}) is also invariant under combined action of coordinate reflection and time
reversal (PT), $x \rightarrow -x$ and $t \rightarrow -t$.

The nonlinear oscillator (\ref{ham1}) admits general periodic solution of the form
\begin{eqnarray}
\fl \qquad \qquad \;\;\; x(t) = \frac{A \sin(\omega t + \delta)}{1 - \frac{k A}{3 \omega} \cos(\omega t + \delta)},
\quad \qquad 0 \le A < \frac{3 \omega}{k},
\label{solu1}
\end{eqnarray}
where $A,\; \delta$ are arbitrary constants. Note that for $0 \le A < \frac{3 \omega}{k}$
(while $-\infty < x < \infty$), the system (\ref{mee}) admits isochronous oscillations of frequency $\omega$,
which is the same as that of the linear harmonic oscillator.  For $A \ge \frac{3 \omega}{k}$, the
solution  becomes singular whenever the phase $(\omega t + \delta)$ takes
the value $\cos^{-1}\left(\frac{3 \omega}{ k A}\right) + 2 n \pi,\; n:$ any integer,
even though $x(t)$ given by
the function in (\ref{solu1}) is periodic of period $T = \frac{2 \pi}{\omega}$, while the corresponding momentum $p$ is
bounded and periodic (see equation (\ref{mom1}) below).  For more details  on the classical dynamics of
this system one may refer to Ref. \cite{chand1}.

Exactly solvable quantum mechanical problems, particularly the ones involving
nonlinear potentials are rare, even in one dimension. The few examples include
P\"{o}schl-Teller, Morse, Scarf and isotonic oscillator potentials \cite{morse}.
Also there exists a few velocity dependent potentials, for example
Mathews-Lakshmanan oscillator and its generalizations \cite{pmm_ml, laksh_higgs}.
The quantization of (\ref{mee}) is a challenging problem since
the obstacles in this task are many. For example, the  quantization of the damped linear harmonic
oscillator itself is a quite complicated procedure requiring a rigged Hilbert space  \cite{damp1} description
whereas the system under consideration is
a nonlinear one. In addition to this, the associated time independent
Hamiltonian is a non-standard one \cite{damp}. To the authors' knowledge
there exists no nonstandard Hamiltonian system  which is quantum
mechanically exactly solvable. Such a system cannot also be quantized using standard techniques of canonical quantization
\cite{dek, gzyl, book_damp}. In this paper, we completely solve the quantum mechanical problem
of the Li\'{e}nard type oscillator (\ref{mee}), possessing the nonstandard Hamiltonian structure (\ref{ham1}),
by associating it with a position dependent mass Hamiltonian where now the variables $x$ and $p$ are interchanged.
We then consider a general symmetric ordered form of the Hamiltonian proposed by von Roos
\cite{von} and solve the underlying Schr\"{o}dinger equation in the momentum space. Since
allowable choices of symmetric ordering lead to singular/unbounded  solution, we transform the symmetric ordered
Hamiltonian suitably in such a way that the associated Schr\"{o}dinger equation possesses acceptable
eigenfunctions. It is worth noting that the transformed Hamiltonian is now  a non-symmetric
ordered one as well as non-Hermitian. But it admits a real energy spectrum
since the Hamiltonian and the corresponding eigenfunctions are invariant under
the action of PT operation when $-\infty < p < \frac{3 \omega^2}{2 k}$.  Our results reveal
that the eigenvalues of (\ref{ham1}) exactly match with that of the linear harmonic
oscillator, though the eigenfunctions are of a more complicated nature in the
momentum space. The explicit form of the eigenfunctions is also presented.
We also obtain the energy level spectrum through a semiclassical
modified Bohr-Sommerfeld quantization rule for the regular periodic solution (\ref{solu1}) which
agrees with the quantum mechanical results. Additionally, we point out the existence of a negative
energy spectrum in the quantum case corresponding to the sector $p > \frac{3 \omega^2}{2 k}$.

We also note here the interesting fact that while the standard PT symmetric systems considered extensively in
the recent literature \cite{bender_r, Ali, Lang_cav}  all correspond to PT invariant complex potentials
involving complex valued dynamical variables, the
present Hamiltonian system (\ref{ham1}) is PT symmetric and  real where the dynamical variables are also
real. The motivation here is
more of exact quantization of a nonlinear dynamical system. Consequently the analysis
of the corresponding quantum system in the momentum space discussed below  will
also be different in spirit from the modified normalization scheme of complexified PT-symmetric
schemes  \cite{bagchi}.

\section{Semiclassical quantization}
To start with let us consider the semiclassical aspects. To quantize the system semiclassically,
we use the modified Bohr-Sommerfeld quantization rule \cite{mbohr}, namely
\begin{equation}
\fl \qquad \qquad  \qquad \oint p \;dx = (n + \frac{1}{2}) h,
\label{bohr}
\end{equation}
where $h$ is the Plank's constant and $n$ is any nonnegative integer and the integration is carried out
over a closed orbit in the $(x, p)$ space.

Firstly, we determine the energy of the system $E = H$ using the general solution (\ref{solu1}) in
(\ref{ham1}). Plugging the expression (\ref{solu1}) in (\ref{mom}), we obtain
\begin{eqnarray}
\fl \qquad \qquad \qquad p \; = A \omega \cos(\omega t +\delta) \left(1 - \frac{k A}{6 \omega} \cos(\omega t + \delta)\right).
\label{mom1}
\end{eqnarray}
Note that for regular (non-singular bounded) periodic oscillations in $x(t)$ given
by equation (\ref{solu1}), the amplitude is
restricted to the range  $0 \leq A < \frac{3 \omega}{k}$, while
the range of $p$ is restricted to $-\frac{9 \omega^2}{2 k} < p < \frac{3 \omega^2}{2 k}$.
As we mentioned earlier (below (\ref{solu1}))
when $A \ge \frac{3 \omega}{k}$, one has singular periodic solution and there is no lower bound on $p$.
Substituting now the expressions (\ref{solu1}) and (\ref{mom1})  in (\ref{ham1}),
we find that the energy of the system turns out to be
\begin{equation}
\fl \qquad \qquad \qquad \qquad  E \;= \frac{1}{2} A^2 \omega^2.
\label{ham2}
\end{equation}

Now, to evaluate the integral (\ref{bohr}), we use (\ref{mom1}) for $p$, and express $dx$ from
(\ref{solu1}) in the form
\begin{eqnarray}
\fl \qquad \qquad \qquad dx \;= \frac{\left(A \cos{\phi} - \frac{k A^2}{3 \omega^2}\right)}{\left(1 - \frac{k A}{3 \omega^2} \cos{\phi}\right)^2} d\phi, \quad
\qquad \phi = \omega t + \delta,
\label{dx}
\end{eqnarray}
so as to obtain the quantization condition for the regular periodic orbits,
\begin{eqnarray}
\fl \qquad \qquad \qquad \omega A^2 \int^{2 \pi}_{0} \left [ \frac{\left(\cos{\phi} - \frac{k A}{6 \omega}\cos^2{\phi}\right)\left(\cos{\phi}
- \frac{k A}{ 3 \omega} \right)}{\left(1 - \frac{k A}{3 \omega}\cos{\phi}\right)^2}\right] d\phi =  (n+\frac{1}{2}) h.
\end{eqnarray}
Evaluating the above integral, we arrive at
\begin{equation}
\fl \qquad \qquad \qquad A^2 \omega = 2 \left(n + \frac{1}{2}\right) \hbar, \quad \quad 0 \le A < \frac{3 \omega}{k}.
\label{bohr2}
\end{equation}

Finally, from equations (\ref{ham2}) and (\ref{bohr2}), we obtain the allowed energy
eigenvalues with an appropriate upper bound $N$ on $n$ corresponding to the regular periodic orbits
of (\ref{ham1}) as
\begin{equation}
\quad E_{n} =  \left( n + \frac{1}{2}\right) \hbar \omega , \qquad n = 0, 1, 2, ... N
\label{energy}
\end{equation}
which agrees with that of the linear harmonic oscillator for this part of the spectrum.
The semiclassical approach motivates us to prove that the energy of the nonlinear oscillator (\ref{ham1})
can be quantized exactly as given in equation (\ref{energy}). In the following, we proceed to solve
the time indepedent Schr\"{o}dinger equation associated with the system (\ref{ham1}) analytically, not in the
coordinate space but in the momentum space.

\section{Quantum exact solvability}
Next we  observe that the classical Hamiltonian $H(x, p)$ given in (\ref{ham1}) is of the non-standard type,
that is,
\begin{eqnarray}
\fl \qquad \quad \qquad \qquad \qquad   H(x, p) = \frac{1}{2} f(p) x^2 + U(p),
\label{pdms_ham}
\end{eqnarray}
where
\begin{eqnarray}
\;\;f(p) = \omega^2\;\left(1 - \frac{2 k }{3 \omega^2} p\right),
\qquad U(p) = \frac{9\omega^4}{2 k^2} \left(\sqrt{1 - \frac{2 k}{3 \omega^2} p}-1\right)^2.
\end{eqnarray}

\begin{figure}[t!]
\centering
\includegraphics[width=0.4\linewidth]{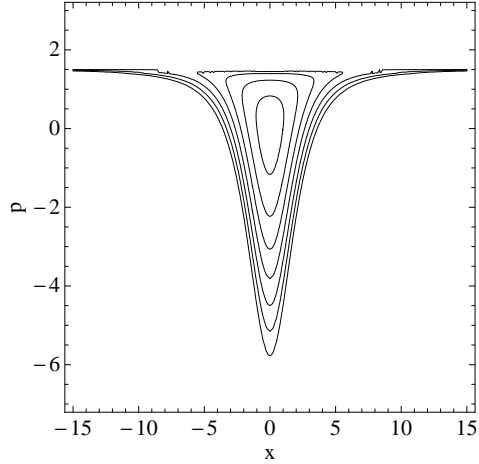}
\vspace{-0.1cm}
\caption{The phase trajectories of the Hamiltonian system (\ref{pdms_ham}) with
$\omega = k = 1$ for various values of $E = H$. }
\label{fig1}
\end{figure}

The ($x-p$) phase space structure is shown schematically in figure \ref{fig1}.
Note the deformed nature of the bounded periodic orbits around the origin and
$-\frac{9 \omega^2}{2 k} < p < \frac{3 \omega^2}{2 k}$.  The remaining trajectories have
only an upper bound at $p = \frac{3\omega^2}{2 k}$.

Note that in the limit $k \rightarrow 0$, $f(p) \rightarrow \omega^2, U(p) \rightarrow \frac{p^2}{2},$
so that
\begin{equation}
\qquad \qquad H = \frac{p^2}{2} + \frac{\omega^2 x^2}{2}
\label{tsho}
\end{equation}
as it should be.  Now the first term in the Li\'{e}nard oscillator Hamiltonian (\ref{pdms_ham})
contains both the position and momentum variables while the second term turns out
to be a function of momentum alone. To quantize the Hamiltonian of this nature, one
has to adopt a suitable ordering procedure. Since $x$ and $p$ are non-commuting variables in the
quantum case, one may consider different ways of ordering between $x$ and $f(p)$ in order to quantize this
Hamiltonian. After performing a detailed analysis we find that the nonstandard
classical Hamiltonian given in (\ref{ham1}) can also be equivalently considered in the form
\begin{eqnarray}
 H(x, p) = \frac{x^2}{2\;m(p)} + U(p), \qquad -\infty < p \le \frac{3\omega^2}{2 k},
\label{pdms}
\end{eqnarray}
where
\begin{equation}
\;\;\;m(p) = \frac{1}{\omega^2\;\left(1 - \frac{2 k }{3 \omega^2} p\right)}
\;\;\;\mbox{and}\;\;\;U(p) = \frac{9\omega^4}{2 k^2} \left(\sqrt{1 - \frac{2 k}{3 \omega^2} p}-1\right)^2.
\label{mass}
\end{equation}
Interestingly, this form is similar to a position dependent mass Hamiltonian,
$H = \frac{p^2}{2\;m(x)} + V(x)$, discussed extensively recently \cite{von, pdms, gonul, ben}
but with an important difference that the variables $x$ and $p$ are now interchanged. Once this fact is
recognized one can consider a general symmetric ordered form of
the quantum Hamiltonian proposed by von Roos
in order to quantize the position (but now actually momentum) dependent mass Schr\"{o}dinger equation \cite{von} as
\begin{eqnarray}
\fl \qquad  \quad \quad H(\hat{x}, \hat{p}) = \frac{1}{4} \left[m^{\alpha}(\hat{p})\hat{x}m^{\beta}(\hat{p})\hat{x}m^{\gamma}(\hat{p}) +
m^{\gamma}(\hat{p})\hat{x}m^{\beta}(\hat{p})\hat{x}m^{\alpha}(\hat{p})\right] + U(\hat{p}),
\label{von}
\end{eqnarray}
where the parameters $\alpha, \beta$ and $\gamma$ which remain to be fixed have to satisfy the condition
 $\alpha + \beta + \gamma = -1$.  Obviously now in (\ref{von}) $\hat{x}$ and $\hat{p}$ are linear
Hermitian operators satisfying the commutation rule
\begin{eqnarray}
\qquad \qquad \qquad[\hat{x}, \hat{p}] = i \hbar.
\end{eqnarray}

We note here that since the variables $\hat{x}$ and $\hat{p}$ are interchanged we solve the Schr\"{o}dinger
equation corresponding to the Hamiltonian (\ref{von}) in momentum space with
${\displaystyle \hat{x} = i \hbar \frac{\partial}{\partial p}}$ and obtain the
time independent one dimensional Schr\"{o}dinger equation in the form
\begin{equation}
\qquad \qquad H(\hat{x}, \hat{p}) \psi(x, p) = E \psi(x, p),
\end{equation}
or
\begin{eqnarray}
\fl  \frac{-\hbar^2}{2 m}\left[\psi^{''} - \frac{m^{'}}{m} \psi^{'} +
\left(\frac{1+\beta}{2}\right) \left(2\frac{m'^2}{m^2} - \frac{m''}{m}\right) \psi +
 \frac{\alpha(\alpha + \beta + 1){m^{'}}^2}{m^2} \psi \right] + U(p)\psi =  E \psi, \hspace{1cm}
 \label{se_a}
\end{eqnarray}
where  prime stands for differentiation with respect to $p$.
Since we are looking for bound states (even when $U(p)$ becomes complex for
$p > \frac{3 \omega^2}{2 k}$),
we can choose
\begin{equation}
\qquad \qquad \qquad  \psi = 0, \;\;\;\mbox{for}\;\;\; p \ge \frac{3 \omega^2}{2 k}
\end{equation}
and concentrate  on the region $-\infty < p \le \frac{3 \omega^2}{2 k}$ alone in this section.
We will also impose  the boundary conditions
\begin{equation}
\qquad \qquad \qquad \psi(-\infty) = \psi\left(\frac{3 \omega^2}{2 k}\right) = 0
\label{bon}
\end{equation}
on the eigenfunctions for continuity and boundedness.

Substituting now the expression for
$m(p)$ from (\ref{mass}) and its derivatives in equation (\ref{se_a}) and simplifying the resultant
expression, we arrive at
\begin{eqnarray}
\fl \;\;\psi^{''} - \frac{2 k}{3 \omega^2} \frac{1}{\left(1 - \frac{2 k}{3 \omega^2} p\right)} \psi^{'}
+ \frac{4 k^2 \alpha(\alpha+\beta+1)}{9\omega^4\;\left(1 - \frac{2k}{3\omega^2} p\right)^2} \psi
- \frac{2 E k^2 - 9\omega^4\left( \sqrt{1 - \frac{2k}{3\omega^2} p}-1\right)^2}{\hbar^2 \omega^2 k^2\left(1 - \frac{2k}{3\omega^2} p\right)} \psi \nonumber \\
\hspace{5cm}= 0, \qquad -\infty < p \le \frac{3 \omega^2}{2 k}.
\label{se_1}
\end{eqnarray}

Equation (\ref{se_1}) can be further  simplified by introducing a transformation
${\displaystyle y^{2} = \left(1 - \frac{2k}{3\omega^2} p\right)}$ or
${\displaystyle y = \sqrt{1 - \frac{2k}{3\omega^2} p}}$. Since
$-\infty < p \le \frac{3 \omega^2}{2 k}$, we have $0 \le y < \infty$.  The resulting 
simplification of (\ref{se_1}) yields
\begin{eqnarray}
\fl \qquad \frac{d^2 \psi}{d y^2} + \frac{1}{y} \frac{d \psi}{dy} + \left( \tilde{E} + \frac{4\alpha(\alpha+ \beta + 1)}{y^2}- a (y - 1)^2 \right) \psi = 0, \quad  0 \le y < \infty,
\label{se1}
\end{eqnarray}
where we have defined ${\displaystyle \frac{18 \omega^2}{\hbar^2 \; k^2} E = \tilde{E}}$,  and
${\displaystyle \frac{3^4 \omega^6}{\hbar^2\;k^4}\;= a}$.

In order to solve (\ref{se1}) subject to the boundary conditions corresponding to (\ref{bon}),
namely $\psi(-\infty) = \psi(0) = 0$, we first note the admissible asymptotic behaviour,
$\psi(y) \rightarrow  e^{-\frac{\sqrt{a}}{2}\left( y^2 - 2 y\right)}$ as $y  \rightarrow \infty$.
So we introduce another transformation, namely
\begin{equation}
\qquad \qquad \qquad \psi(y) = e^{-\frac{\sqrt{a}}{2} y^2 + \sqrt{a} y} \; \phi(y)
\label{trans1}
\end{equation}
 in (\ref{se1}) so that it can be rewritten as
\begin{eqnarray}
\fl \qquad \;\; \frac{d^2 \phi}{dy^2} + \left(2 \sqrt{a} - 2 \sqrt{a} y + \frac{1}{y}\right) \frac{d \phi}{d y}
+ \left(\frac{4\alpha(\alpha+ \beta + 1)}{y^2} + \frac{\sqrt{a}}{y} + \tilde{E} - 2 \sqrt{a} \right)\phi = 0.
\label{se3}
\end{eqnarray}
Equation (\ref{se3}) can  now be transformed to  the Hermite differential equation under the change of
variables
\begin{eqnarray}
\qquad \qquad  \phi(y) = y^{-1/2} \chi(z), \qquad z = a^{1/4}(y -1),
\label{phi}
\end{eqnarray}
with the condition $4 \alpha (\alpha + \beta + 1) = -\frac{1}{4}$. The transformed equation
turns out to be of the form
\begin{eqnarray}
\qquad \quad \frac{d^2 \chi}{dz^2} - 2z\frac{d\chi}{dz} + \left(\frac{\tilde{E} - \sqrt{a}}{\sqrt{a}}\right) \chi = 0.
\label{se4}
\end{eqnarray}
With the restriction of  the constant ${\displaystyle \frac{\tilde{E} - \sqrt{a}}{\sqrt{a}} = 2 n}$,
$n = 0, 1, 2, ...,$ equation (\ref{se4}) becomes the standard differential equation for
the Hermite polynomials,
\begin{eqnarray}
\qquad \quad \frac{d^2 \chi}{dz^2} - 2z\frac{d\chi}{dz} + 2 n \chi = 0, \qquad n = 0, 1, 2, ...
\label{hermite}
\end{eqnarray}
where $\chi = H_{n}(z)$ which are nothing but the Hermite polynomials \cite{book}.
Then the eigenfunctions and eigenvalues can  be  readily written down as
\begin{eqnarray}
\fl \qquad \psi_n(p) = N_n \frac{\exp{\left(-\frac{9\omega^{3}}{2\;\hbar\;k^2}\left(1 - \frac{2k}{3\omega^2}p - 2\sqrt{1 - \frac{2k}{3\omega^2}p}\right)\right)}}{\left(1 - \frac{2k}{3\omega^2}p\right)^{1/4}} \nonumber \\
\qquad \qquad \times H_{n}\left[\frac{3\omega^{3/2}}{\sqrt{\hbar}\;k}\left(\sqrt{1-\frac{2k}{3\omega^2}p}- 1\right)\right], \; - \infty < p \le \frac{3 \omega^2}{2 k},
\label{eig}
\end{eqnarray}
\begin{eqnarray}
\fl  \qquad \qquad &=& 0, \qquad \quad  \frac{3 \omega^2}{2 k} \le p < \infty,
\label{eig_1}
\end{eqnarray}
and
\begin{eqnarray}
\fl \qquad \qquad  \quad E_{n} &=& (n+\frac{1}{2}) \hbar \omega, \qquad \quad n = 0, 1, 2, ... .
\label{sol}
\end{eqnarray}
Here $N_n$ are constants.

Note that the above eigenfunction is singular at the boundary $p = \frac{3 \omega^2}{2 k}$ due
to the denominator term in the right hand side of the equation (\ref{eig}) and so
the eigenfunction becomes unbounded. To avoid this singularity in the eigenfunction,
we modify the starting Hamiltonian, $H(\hat{x}, \hat{p})$, suitably
and solve the associated Schr\"{o}dinger equation.  For this purpose we can
rewrite the time independent Schr\"{o}dinger equation $H\psi = E \psi$ as
\begin{equation}
\qquad \quad \quad  H (m^{-d} \Phi) = E (m^{-d} \Phi),
\label{mham}
\end{equation}
so that we have
\begin{equation}
\fl \qquad \qquad \tilde{H}\;\Phi = (m^{d} H m^{-d})\;\Phi = E \Phi \;\;\; \mbox{and} \;\;\;
\Phi = m^{d}\;\psi, \;\;\; m(p) = \frac{1}{\omega^2
\left(1 - \frac{2 k}{3 \omega^2} p\right)}.
\label{mham}
\end{equation}
With the choice $d <-\frac{1}{4}$, one can have bounded, continuous and single valued wavefunction
for the Hamiltonian $\tilde{H}$.

As a simple choice we consider a specific set of the values of the ordering
parameters $\alpha = \gamma = -\frac{1}{4}$ and
$\beta = -\frac{1}{2}$ that satisfy the conditions $\alpha + \beta + \gamma = -1$ and
$4\alpha(\alpha+\beta+1) = -\frac{1}{4}$ so that $d = -\frac{1}{2}$. Hence the Hamiltonian $H(\hat{x}, \hat{p})$ in (\ref{von}) becomes
\begin{eqnarray}
\fl \qquad  \quad \quad H(\hat{x}, \hat{p}) = \frac{1}{2} \left[m^{-1/4}(\hat{p})\hat{x}m^{-1/2}(\hat{p})\hat{x}m^{-1/4}(\hat{p})\right] + U(\hat{p}),
\label{von2}
\end{eqnarray}
which admits the solution as given in (\ref{eig}). Now we transform the Hamiltonian, given in
(\ref{von2}), to the form
\begin{eqnarray}
\fl \; \tilde{H} &=& \frac{1}{\sqrt{m}} H \sqrt{m} = \frac{1}{2} \left[m^{-3/4}(\hat{p})\hat{x}m^{-1/2}(\hat{p})\hat{x}m^{1/4}(\hat{p})\right] + U(\hat{p}), \; \label{t_ham}\\
\fl \;           &=& -\frac{\hbar^2}{2} \omega^2 \left(1 - \frac{2 k}{3 \omega^2} p \right) \left[ \frac{d^2}{d p^2} + \frac{k^2}{ 12 \omega^4} \frac{1}{\left(1 - \frac{2 k}{3 \omega^2} p \right)^2}\right] + \frac{9 \omega^4}{2 k^2} \left(\sqrt{1 - \frac{2 k}{3 \omega^2} p }  - 1\right)^2. \qquad
\label{nham}
\end{eqnarray}
This Hamiltonian (\ref{nham}) is invariant under PT symmetry \cite{bend} though it is nonsymmetric and non-Hermitian.
The Schr\"{o}dinger equation corresponding to the Hamiltonian, $\tilde{H}(x, p)$  is
\begin{eqnarray}
\fl \quad -\frac{\hbar^2 \omega^2 }{2}\left(1 - \frac{2 k}{3 \omega^2} p \right) \Phi'' - \frac{\hbar^2 k^2  }{ 24 \omega^2\left(1 - \frac{2 k}{3 \omega^2} p\right)} \Phi+ \frac{9 \omega^4}{2 k^2} \left(
\sqrt{1 - \frac{2 k}{3 \omega^2}p} - 1\right)^2 \Phi  = E \Phi, \;\;\left('= \frac{d}{d p}\right). \qquad
\label{seh}
\end{eqnarray}
Equation (\ref{seh}) is now solved again by following the above procedure. The resulting 
bound state solution turns out to be
\begin{equation}
\fl \quad \Phi_n(p) = \left\{\begin{array}{cc} \tilde{N}_n \left(1 - \frac{2k}{3\omega^2}p\right)^{1/4} \exp{\left(-\frac{9\omega^{3}}{2\;\hbar\;k^2}\left(1 - \frac{2k}{3\omega^2}p - 2\sqrt{1 - \frac{2k}{3\omega^2}p}\right)\right)}\nonumber \\
\hspace{3cm} \times H_{n}\left[\frac{3\omega^{3/2}}{\sqrt{\hbar}\;k}\left(\sqrt{1-\frac{2k}{3\omega^2}p} - 1\right)\right],
-\infty < p \le \frac{3 \omega^2}{2 k}, \label{sol2a} \\
                                  \hspace{-7cm}0, \quad p \ge \frac{3 \omega^2}{2 k},
                                   \end{array}\right.
\label{sol2}
\end{equation}
and the corresponding energy eigenvalues continue to be
\begin{eqnarray}
\fl \hspace{4cm}  E_{n} &=& (n+\frac{1}{2})\;\hbar\; \omega, \qquad \quad n = 0, 1, 2, ....
\label{sol1}
\end{eqnarray}
One can observe that  the solution (\ref{sol2}) is continuous, single valued and bounded in
the entire region $-\infty < p < \infty$ and satisfy the boundary conditions
$\Phi(\infty) = \Phi(3k/2\omega^2) = \Phi(-\infty) = 0$. Since the eigenfunctions $\Phi_n(p)$ are bounded and continuous in the region
$-\infty < p \le \frac{3 \omega^2}{2 k}$ and are zero outside this region, they are also normalizable. The normalization
constants $\tilde{N}_n$ can be found using the integration
\begin{eqnarray}
\int^{\frac{3\omega^2}{2k}}_{-\infty} \Phi^{*}_{n}(p)\; \Phi_{n}(p) dp = 1.
\end{eqnarray}
This can be evaluated as
\begin{eqnarray}
\int^{0}_{-\infty} \Phi^{*}_{n}(p)\; \Phi_{n}(p) dp + \int^{\frac{3\omega^2}{2k}}_{0} \Phi^{*}_{n}(p)\; \Phi_{n}(p) dp = 1.
\label{n1}
\end{eqnarray}
On evaluating (\ref{n1}) becomes
\begin{eqnarray}
\tilde{N}^2_n\sqrt{\hbar \omega}\; e^{\left(\frac{9 \omega^{3}}{k^2 \hbar}\right)}\; \left( 2^{n -1} \sqrt{\pi} n! \left(1 + \frac{9 \omega^{3}}{k^2 \hbar}\right) + g(a)\right) = 1,
\label{norm1}
\end{eqnarray}
where
\begin{eqnarray}
g(a) =  \int^{\frac{3\omega^2}{2k}}_{0} \Phi^{*}_{n}(p)\; \Phi_{n}(p) dp.
\end{eqnarray}
Hence the normalization constant is
\begin{eqnarray}
\tilde{N}_n = \left(\frac{e^{-\left(\frac{9 \omega^{3}}{k^2 \hbar}\right)}}{\sqrt{\hbar \omega}(2^{n-1} \sqrt{\pi} n! \left(1 + \frac{9 \omega^{3}}{k^2 \hbar}\right) + g(a))} \right)^{1/2}.
\label{norm2}
\end{eqnarray}
We further note that in the limit $k \rightarrow 0$,
equation (\ref{sol2a}) reduces to
\begin{eqnarray}
\fl \qquad \qquad \Phi_n(p) = \left(\frac{-1}{2^n \sqrt{\pi}\; n!} \right)^{1/2} \exp{\left(-\frac{1}{2\;\hbar\;\omega} p^2 \right)} H_{n}\left[\frac{1}{\sqrt{\hbar \omega}} p\right], \;\;\;
-\infty < p < \infty,
\label{sho1}
\end{eqnarray}
which matches with the bound state solution of the harmonic oscillator in accordance with its classical
counter part (vide (\ref{tsho})).

Finally one can also note that one can choose many number of possible non-symmetric Hamiltonian $\tilde{H}$ in
(\ref{t_ham}) for suitable choice of the set of parameters $\alpha, \beta, \gamma$ and $d$ all of which lead
to the same eigenvalue spectrum but different sets of eigenfunctions.

\section{The $p> \frac{3\omega^2}{2 k}$ sector: broken symmetry}

In addition to the above solutions, we can also identify a different set
of solutions which is nonzero only in the  regime $p > \frac{3 \omega^2}{2k}$ with a
different set of boundary conditions than (\ref{bon}). To realize this, we
consider solutions with $\Phi = 0$, $-\infty < p \le \frac{3 \omega^2}{2 k}$, and
look for acceptable solutions in the region $\frac{3 \omega^2}{2 k} < p \le \infty$, either bound
states with the boundary condition $\Phi(\frac{3 \omega^2}{2 k}) = 0 = \Phi(\infty)$ or
continuum states or both. We also note that in the case of the classical nonlinear
oscillator (\ref{mee}) with the Hamiltonian (\ref{ham1}), there exists no real solution for $p > \frac{3 \omega^2}{2 k}$
due to the form of the conjugate momentum (\ref{mom}).

For the above purpose, we consider the Schr\"{o}dinger equation (\ref{seh})
corresponding to the Hamiltonian (\ref{nham}), under the
transformation $ \tilde{y}  = \sqrt{\frac{2 k}{3 \omega^2}p - 1}$, as
\begin{eqnarray}
\fl \qquad \frac{d^2 \Phi}{d \tilde{y}^2} - \frac{1}{\tilde{y}} \frac{d \Phi}{d\tilde{y}} + \left(- \tilde{E} + \frac{3}{4\tilde{y}^2}
+a (i \tilde{y} - 1)^2 \right) \Phi = 0, \quad  0 \le \tilde{y} < \infty,
\label{se11}
\end{eqnarray}
where again  we have defined ${\displaystyle \frac{18 \omega^2}{\hbar^2 \; k^2} E = \tilde{E}}$,  and
${\displaystyle \frac{3^4 \omega^6}{\hbar^2\;k^4}\;= a}$.
Under the transformation ${\displaystyle \Phi(\tilde{y})} = \sqrt{\tilde{y}} e^{-\frac{\sqrt{a} }{2}\tilde{y}^2  - i \sqrt{a} \tilde{y}} \chi(z)$
with $ z = a^{1/4} (\tilde{y} + i)$, we get
\begin{eqnarray}
\frac{d^2 \chi(z)}{d z^2} - 2 z \frac{d \chi(z)}{dz} - \frac{\tilde{E} + \sqrt{a}}{ \sqrt{a}}\chi(z) = 0,
\label{se_f}
\end{eqnarray}
so that the bounded solution (as $|z| \rightarrow \infty \; or \; \tilde{y} \rightarrow \infty$) can be now expressed
(compared to (\ref{se4})) as
\begin{equation}
\chi(z) = H_{n}(z), \quad  \tilde{E} = \frac{18 \omega^2}{\hbar^2 k^2} E = -(2n + 1) \sqrt{a}, \quad n = 0, 1, 2, ... .
\end{equation}
Thus the second set of solution to the Schr\"{o}dinger equation (\ref{seh}) can be
written as
\begin{equation}
\fl \quad \Phi_n(p) = \left\{\begin{array}{cc} \tilde{\cal{N}}_n \left(\frac{2k}{3\omega^2}p - 1\right)^{1/4} \exp{\left(-\frac{9\omega^{3}}{2\;\hbar\;k^2}\left(\frac{2k}{3\omega^2}p -1 + i \;2\sqrt{\frac{2k}{3\omega^2}p -1}\right)\right)}\nonumber \\
\hspace{3cm} \times H_{n}\left[\frac{3\omega^{3/2}}{\sqrt{\hbar}\;k}\left(\sqrt{\frac{2k}{3\omega^2}p - 1} + i\right)\right],
              \quad p \ge \frac{3 \omega^2}{2 k}, \label{eig2a} \\
                                  \hspace{-7cm}0, \quad -\infty < p \le \frac{3 \omega^2}{2 k},
                                   \end{array}\right.,
\label{sol3}
\end{equation}
with energy eigenvalues without a lower bound
\begin{equation}
E_n = - (n + \frac{1}{2}) \hbar \omega, \quad n = 0, 1, 2, 3, ... .
\end{equation}
In (\ref{sol3}) $\tilde{\cal{N}}_n$ is the normalization constant.
Note that the eigenfunctions (\ref{sol3}) are no longer PT symmetric, even though the Hamiltonian (\ref{nham})
 is PT symmetric, leading to a negative energy spectrum that is unbounded below. Such
 a broken symmetry is obviously a consequence of imposition of a different set of boundary
 conditions for the sector $p > \frac{3 \omega^2}{2 k}$ than (\ref{bon}) which is
 reminiscent of the situation in the case of the linear harmonic oscillator \cite{bender_r, bend_neg}
 and the $H = \hat{p}^2 - \hat{x}^4$ oscillator \cite{bender_r}.

\section{Conclusion}
We have shown that the non-Hermitian Hamiltonian $\tilde{H}(\hat{x}, \hat{p})$ given by
(\ref{t_ham}) admits the bound state solutions, $\Phi_n(p)$ and real energy eigen spectrum, $E_n$.
The energy eigenvalues $E_n$ also match with the energy values obtained through a
semiclassical approach corresponding to regular periodic orbits. It is interesting to
observe that the quantum system (\ref{von2}) possesses the
energy eigenvalues $E_n$ which are same as that of the linear harmonic oscillator, though the eigenfunctions are
quite different from that of the linear harmonic oscillator. Our analysis shows that the underlying
Li\'{e}nard type PT-invariant reversible nonlinear oscillator is exactly quantizable and leads to interesting class
of eigenfunctions and energy spectrum. It is also possible to generalize the above results to
more general class of Li\'{e}nard type nonlinear oscillators \cite{cha} and
coupled nonlinear oscillators \cite{couple}, which will be taken up elsewhere.

\section{Acknowledgments}
VC wishes to thank the Council of Scientific and Industrial Research,
Government of India, for providing a Senior Research Fellowship. The work forms
a part of a research project of MS, and an IRHPA project and a Ramanna Fellowship project
of ML, sponsored by the Department of Science and Technology (DST), Government of India. ML
also acknowledges the financial support under a DAE Raja Ramanna Fellowship.


\section*{References}
\begin{thebibliography}{11}

\bibitem{chand1}
Chandrasekar V K, Senthilvelan M and Lakshmanan M 2005 \emph{Phys. Rev. E} {\bf 72} 066203

\bibitem{bend}
Bender C M and Boettcher S 1998 \emph{Phys. Rev. Lett.} {\bf 30} 5243;
Fring A 2007 \emph{Acta Polytech.} 47 44

\bibitem{morse}
Belchev B and Walton M A 2010 {\it The Morse potential and phase-space quantum mechanics}
arXiv.org:1001.4816v1; Zhu D 1987 \emph{J. Phys. A: Math. Gen.} {\bf 20} 4331

\bibitem{pmm_ml}
Mathews P M and Lakshmanan M 1975 \emph{Nuovo Cimento A} {\bf 26} 299; 1974 \emph{Q. Appl. Math.} {\bf 32} 215;
Midya B and Roy B 2009 \emph{J. Phys. A: Math. Theor.} {\bf 42} 285301

\bibitem{laksh_higgs}
Lakshmanan M and Eswaran K  1975 \emph{J. Phys. A: Math. Gen.} {\bf 8} 1658;
Higgs P W 1979 \emph{J. Phys. A: Math. Gen.} {\bf 12} 309

\bibitem{damp1}
Chandrasekar V K, Senthilvelan M and Lakshmanan M 2007 \emph{J. Math. Phys.} {\bf 48}, 032701

\bibitem{damp}
Gladwin Pradeep R, Chandrasekar V K, Senthilvelan M and Lakshmanan M 2009 \emph{J. Math. Phys.} {\bf 50} 052901


\bibitem{dek}
Dekker H 1977  \emph{Phys. Rev. A} {\bf 16} 2126

\bibitem{gzyl}
 Gzyl H  1983 \emph{Phys. Rev. A} {\bf 27} 2297

\bibitem{book_damp}
Razavy M 2005 {\it Classical and quantum dissipative systems} (London: Imperial College Press)


\bibitem{von}
von Roos O 1983 \emph{Phys. Rev. B} {\bf 27} 7547; 
von Roos O  and Mavromatis H 1985 \emph{Phys. Rev. B} {\bf 31} 2294

\bibitem{bender_r}
Bender C M and Hook D W 2011 \emph{J. Phys. A: Math. Theor.} {\bf 44} 372001;
Bender C M 2007 \emph{Rep. Prog. Phys.} {\bf 70} 947

\bibitem{Ali}
Mostafazadeh A 2010 \emph{Int. J. Geom. Meth. Mod. Phys.} {\bf 7} 1191

\bibitem{Lang_cav}
Langer H and Tretter C 2004 \emph{Czech. J. Phys.} {\bf 54} 1113;
Cavaglia A, Fring A and Bagchi B 2011 \emph{J. Phys. A: Math. Theor.} {\bf 44} 325201

\bibitem{bagchi}
Bagchi B, Quesne C, Znojil M 2001 \emph{Mod. Phys. Lett.A} {\bf 16} 2047


\bibitem{mbohr}
Marinov M S and  Popov V S 1975 \emph{J. Phys. A: Math. Gen.} {\bf  8} 1575


\bibitem{pdms}
Bastard G 1992 \emph{Wave Mechanics Applied to Semiconductor Heterostructures} (Les Ulis: Les  Editions de Physique)

\bibitem{gonul}
G\"{o}n\"{u}l B, \"{O}zer O, G\"{o}n\"{u}l B and \"{U}zg\"{u}n F 2002 \emph{Mod. Phys.  Lett. A} {\bf 17}  2453

\bibitem{ben}
Koc R and Koca M 2003 \emph{J. Phys. A: Math. Gen.} {\bf 36} 8105;
Ju G-X, Cai C-Y and Ren Z-Z  2009 \emph{Commun. Theor. Phys.} {\bf 51} 797;

\bibitem{book}
Brychkov Y A 2008 {\it Handbook of special functions: derivatives, integral series and other formulas} 
(Boca Raton, FL: Chapaman and Hall/CRC)


\bibitem{bend_neg}
Bender C M, Hook D W and  Klevansky S P 2012 {\it Negative-energy PT-symmetric Hamiltonians}
 arXiv:1203.6590

\bibitem{cha}
Gladwin Pradeep R, Chandrasekar V K, Senthilvelan M and Lakshmanan M 2010 \emph{J. Math. Phys.} {\bf 51} 033519

\bibitem{couple}
Chandrasekar V K, Senthilvelan M and Lakshmanan M 2005 \emph{Proc. R. Soc. A} {\bf 461} 2451;
Gladwin Pradeep R, Chandrasekar V K, Senthilvelan M and Lakshmanan M 2009 \emph{J. Phys. A: Math. Theor.} {\bf 42} 135206


\end{thebibliography}
\end{document}